\documentclass[12pt]{article}
\usepackage{epsfig}
\usepackage{amssymb}

\usepackage{color}

\newtheorem{Lemma}{Lemma}

\newcommand{\qed}{$\Box$}
\renewcommand{\rho}{\varrho}
\newcommand{\dx}{\,\mathrm{d}x}
\newcommand{\pat}{\partial_t}

\newcommand{\nn}{\nonumber}
\newcommand{\half}{\frac{1}{2}}
\newcommand{\ve}{\varepsilon}
\newcommand{\tr}{\mathrm{tr}}
\newcommand{\io}{\int\limits_\Omega}
\newcommand{\R}{\mathbb R}

\newcommand{\SO}{\mathrm{SO}}
\newcommand{\sym}{\mathrm{sym}}
\newcommand{\skw}{\mathrm{skw}}

\newcommand{\Fe}{F_\mathrm{e}}
\newcommand{\Fp}{F_\mathrm{p}}
\newcommand{\Ue}{U_\mathrm{e}}
\renewcommand{\Re}{R_\mathrm{e}}
\newcommand{\Ke}{K_\mathrm{e}}
\newcommand{\Fpo}{F_{\mathrm{p}_0}}
\newcommand{\Wst}{W_{\mathrm{st}}}
\newcommand{\Wc}{W_{\mathrm{c}}}
\newcommand{\Id}{\mathrm{Id}}
\renewcommand{\phi}{\varphi}
\newcommand{\fext}{f_\mathrm{ext}}
\newcommand{\Mext}{M_\mathrm{ext}}
\newcommand{\aopt}{\alpha_\mathrm{opt}}
\newcommand{\Ad}{\dot{A}}

\newcommand{\kd}{\dot{k}}
\newcommand{\gd}{\dot{\gamma}}

\begin{document}
\author{T.~Blesgen (blesgen@mis.mpg.de)\\
Max-Planck-Institute for Mathematics in the Sciences,\\
Inselstra{\ss}e 22-26, D-04103 Leipzig, Germany}

\title{Domain partitioning as a
result of deformation in the framework of large-strain Cosserat plasticity}
\maketitle

\begin{abstract}
In the framework of the rate-independent large-strain Cosserat theory of
plasticity we calculate analytically explicit solutions of a two-dimensional
shear problem. We discuss two cases where the micro-rotations are stationary
solutions of an Allen-Cahn equation. Thus, for a certain parameter range,
patterning arises and the domain is partitioned into subsets with approximately
constant rotations. This describes a possible mechanism for the formation
of grains and subgrains in deformed solids.
\end{abstract}

\vspace{2pc}

\vspace{2pc}
\hspace*{58pt}Allen-Cahn equation

\vspace{2pc}
\noindent{Submitted to Modelling Simulation Mater. Sci. Eng.}
\newpage

\section{Introduction}
\label{secintro}
Experimental evidence suggests that the size of grains and subgrains in
plastically deformed solids is not random, but rather relates inversely
to the flow stress,
\[ \sigma\sim D^{-1/2}, \]
see \cite{CGG06, CGG06b}. The foundations of this heuristic law are a
long-standing open problem in materials science and solid mechanics.
The qualitative and quantitative prediction of $D$ is particularly
vexing in dynamic recrystallisation (DRX), where new grains essentially free
of dislocations nucleate out of a highly fatigued sample. The experimental
studies \cite{FG08} on DRX show that the recrystallized grain sizes
can be linked to the subgrain size although the dependency of dynamically
recrystallized grain size and subgrain size on flow stress is different.

Derby and Ashby \cite{DA87}, and Roberts et al., \cite{RA78}, addressed the
grain size by making assumptions on the relevant criteria like back stresses at
grain boundaries or imbalance of subgrain size across prior grain boundaries
but did not touch the issue of mechanical behavior. A survey on the effect of
diffusion on grain boundaries can be found in \cite{Atk88}.

Evidently, despite extensive research during the past decades, the
physical mecha\-nisms of grain formation are not yet understood. The subject is
not only of academic interest but is also of immense importance for industrial
processing, see e.g. \cite{Raabe1,Raabe2,CB00,Ungar01},
since treatments like hot rolling, cold rolling and annealing
require reliable predictions of the microstructure and the texture evolution.

In this article, without spurious assumptions or introducing an artificial
mechanism, we show in a mathematically rigorous way why patterning in
deformed solids may occur. To develop our argument we compute analytically in
special cases the state of a solid that is deformed along its boundary.
We formulate the problem within the finite-strain Cosserat model of plasticity.
The interesting feature of this model, unlike other established approaches in
visco-plasticity, see the research articles and surveys
\cite{Simo,Simo2,Miehe98,Hill98}, is that it is a {\it gradient model}, i.e. a
length scale is introduced. The results given here substantiate this property:
it is demonstrated how for a certain parameter range the analytic solutions
naturally lead to transition layers between areas of approximately constant
micro-rotations $\Re$, a parameter that specifies the local orientation
of the material.

This paper is organised in the following way. In Section~\ref{secmodel} we
revise the rate-independent finite-strain Cosserat model as needed later
and formulate an energy minimisation procedure that allows to obtain the
time-discrete solution. We then restrict the plastic deformations to a priori
given single-slip systems and adopt the formulation accordingly.
In Section~\ref{secspecial}, we apply the resulting model to a simple shear
problem in 2D. The two-dimensional setting is chosen as this permits the
representation of the rotations by one scalar parameter.
In the following section \ref{secsol} some exact solutions to this shear
problem are computed with focus on the properties of the micro-rotations $\Re$.
As main result, two non-trivial examples are discussed where $\Re$ is connected
with the solutions of a stationary Allen-Cahn equation. We end with a discussion
of the results.

\section{The rate-independent finite-$\!$strain Cosserat model of
visco-plasticity}
\label{secmodel}
Let $\Omega\subset\R^d$ be a bounded domain with Lipschitz boundary that
serves as a reference configuration. The deformation of the material is
controlled by a mapping $\phi(\cdot,t)$ that maps $\Omega$ diffeomorphically to
the deformed state $\Omega_t$ at time $t\ge0$. Because of $\phi(\cdot,0)=\Id$
it holds $\det(D\phi(t))>0$ for all $t\ge0$.

The gist of the Cosserat approach is to multiplicatively decompose the
deformation tensor $F:=D\phi$ into a plastic part $\Fp$ and an elastic part
$\Fe$ by virtue of
\begin{equation}
\label{decomp}
F=\Fe\Fp,
\end{equation}
and to split $\Fe$ by
\begin{equation}
\label{decomp2}
\Fe=\Re\Ue
\end{equation}
into a stretching component $\Ue\in\mathrm{GL}(\R^d)$ and a rotation part
$\Re\in\SO(d)$, where
\[ \SO(d):=\{R\in\mathrm{GL}(\R^d)\;|\;\det(R)=1,\,R^tR=\Id\} \]
denotes the special orthogonal group. Equation~(\ref{decomp}) states
that the plastic deformation of the material precedes the elastic deformation.
In general, $\Ue$ need not be symmetric and positive definite, i.e.
(\ref{decomp2}) is \textit{not} the polar decomposition.

Introducing the dislocation density $\kappa$, frame indifference (which
amounts here to the invariance of the energy under rigid body motions)
implies that the mechanical energy density is the sum of three functionals,
\begin{equation}
\label{split}
W(\Fe,\kappa)=\Wst(\Ue)+\Wc(\Ke)+V(\kappa),
\end{equation}
with the stretching part $\Wst$, the curvature part $\Wc$, and the energy due
to immobilised dislocations $V$.
With $\Ke:=\Re^tD_x\Re=(\Re^t\partial_{x_l}\Re)_{1\le l\le d}$ we
designate the third-order (right) curvature tensor. Eqn.~(\ref{split}) tacitly
assumes that the mechanical energy depends on the elastic energy $\Fe$ only.
For arbitrary materials, this need not be the case.

Following the ideas in \cite{OR99}, we now derive a time-discrete formulation
that allows to compute the evolution of the material by minimising the
mechanical energy. For known values $(\Fpo,\kappa_0)$ of the previous time step
and discrete time step $h>0$, the values of $(\phi,\Re,\Fp,\kappa)$ at time
$t+h$ are calculated.

With $P:=\Fp^{-1}$, $P_0:=\Fpo^{-1}$ we choose the discretisations
\[ d_t^h(\Fp):=\frac{\Id-P^{-1}P_0}{h},\qquad
\pat^h\kappa:=\frac{\kappa-\kappa_0}{h}. \]
By $\fext(t)$ we denote external volume forces applied to the crystal body,
$\Mext(t)$ are external volume couples. Let $\Gamma_D\subset\partial\Omega$ be
the Dirichlet boundary (for simplicity postulated invariant in time)
which we assume to be a smooth curve with positive $(d\!-\!1)$-dimensional
Hausdorff measure. On $\Gamma_D$ we prescribe the Dirichlet boundary conditions
\begin{equation}
\label{BC} \phi_{|\Gamma_D}=g_D,\qquad {\Re}_{|\Gamma_D}=R_D.
\end{equation}
Since we are only interested in the traction-free case, we henceforth set
$\Gamma_D:=\partial\Omega$.
The time-discrete mechanical energy functional for given $(\Fpo,\kappa_0)$ then
reads
\begin{eqnarray}
E(\phi,\Re,\Fp,\kappa)(t) \!\!\!&:=&\!\!\! \io\Big[\Wst(\Ue)+\Wc(\Ke)+V(\kappa)
-\fext(t)\!\cdot\!\phi-\Mext(t)\!:\!\Re\nn\\
\label{Edef}
&& \hspace*{25pt} +hQ^*(d_t^h(\Fp),\pat^h\kappa)\Big]\dx.
\end{eqnarray}
The term $hQ^*(d_t^h(\Fp),\pat^h\kappa)$ in (\ref{Edef}) represents
the dissipated mechanical energy in a discrete time interval of length $h$,
defined by the Legendre-Fenchel dual
\begin{equation}
\label{Qstardef}
Q^*(\Fp,\kappa):=\sup_{(X,\pi)}\{X\!:\!\Fp+\pi\kappa-Q(X,\pi)\}
\end{equation}
of the plastic potential
\[ Q(X,\pi):=\left\{\begin{array}{r@{,\quad}l} 0 & \mbox{for }
Y(X,\pi)\le0,\\ \infty & \mbox{else} \end{array}\right. \]
and $Y$ is the yield function
\begin{equation}
\label{Ydef}
Y(\sigma,\pi):=\|\mathrm{dev}\,\sym\,\sigma\|-\sigma_Y-\pi\le0.
\end{equation}
This is the von Mises approach,
$\mathrm{dev}\,\sigma:=\sigma-\frac1d\tr(\sigma)\Id$ designates the
deviatoric part of a tensor $\sigma$, $\sym\,\sigma:=0.5(\sigma+\sigma^t)$,
and $\sigma_Y>0$ is the yield stress.

The solution of the minimisation problem
\begin{equation}
\label{Emin}
E(\phi,\Re,\Fp,\kappa)\to\min
\end{equation}
subject to the boundary conditions (\ref{BC}) yields $(\phi,\Re,\Fp,\kappa)$ at
time $t+h$. The first variation of $E$ w.r.t. the (implicit dual) variables
$(X,\pi)$ leads to the time-discrete plastic flow rule (see \cite{Simo}
for details)
\begin{equation}
\label{flowrule}
(\Fpo d_t^h(\Fp),\pat^h\kappa)\in\partial^\mathrm{sub}Q(X,\pi).
\end{equation}
The variation of $E$ w.r.t. $(\phi,\Re)$ gives back the field equations.
The variation w.r.t. $(\Fp,\kappa)$ returns the definition of the back stress
$X$ and the hardening modulus $\pi$,
\[ X=-\frac{\partial\Wst(\Ue)}{\partial\Fp},\qquad\pi=-V'(\kappa). \]

\section{Single-slip systems}
\label{secss}
We specialise the minimisation problem (\ref{Emin}), (\ref{Edef}) to 
single-slip plasticity. In \cite{CO06}, it had been shown that the relaxed
behavior of crystals restricted to single-slip under the infinite latent
hardening constraint is identical to the dynamics of materials with general
multislip plasticity.

Let $m_a\in\R^d$ be given slip vectors, $n_a\in\R^d$ be
given slip normals with $|m_a|=|n_a|=1$, $m_a\!\cdot\! n_a=0$ for
$1\le a\le I$, $I$ the number of slip systems present. Let
\[ Y(\sigma,\pi):=\max_{1\le a\le I}\Big\{|m_a\!\cdot\!\sigma n_a|\Big\}
-\sigma_Y-\pi \]
be the corresponding yield function. For single-slip systems,
\[ \Fp=\Fp(\gamma):=\Id+\sum_{a=1}^I\gamma_am_a\!\otimes\!n_a, \]
with the parametrisation $\gamma=(\gamma_a)_{1\le a\le I}\in\R^I$.

As can be checked, the dissipated energy satisfies the identity
\[ Q^*(\Ad,\kd)=\left\{\!\!
\begin{array}{l@{\quad}l}\sigma_Y\sum_{a=1}^I|\gd_a|,
& \!\mbox{if }\Ad\!=\!\sum_{a=1}^I\gd_a\,m_a\!\otimes\!n_a\mbox{ and }
\sum_{a=1}^I|\gd_a|+\kd\le0,\\
\infty, & \!\mbox{else.} \end{array}\right. \]
Therefore (\ref{Emin}) becomes
\begin{eqnarray}
E(\phi,\Re,\gamma,\kappa) &=& \io\Big[\Wst(\Re^tD\phi\Fp(\gamma)^{-1})+\Wc(\Ke)
+V(\kappa)-\fext\!\cdot\!\phi\nn\\
\label{Emin2}
&& \hspace*{25pt} -\Mext\!:\!\Re+\sigma_Y\sum_{a=1}^I
|\gamma_a-\gamma_a^0|\Big]\dx\to\min
\end{eqnarray}
subject to
$f_1(\gamma,\kappa):=\sum_{a=1}^I|\gamma_a-\gamma_a^0|+\kappa-\kappa_0\le0$
and (\ref{BC}).

From now on, we set $\Mext=\mathbf{0}$, $\fext=0$.

\section{The shear problem in $2D$}
\label{secspecial}
In general, even in two space dimensions the minimisation problem (\ref{Emin2})
is too complex to be solved analytically and numerical computations are
required. In order to proceed, we make two simplifying assumptions. The first
assumption is that there is only one active slip system, $I=1$, such that
\begin{equation}
\label{Fpdef}
\Fp(\gamma)=\Id+\gamma\,m\!\otimes\!n,\qquad
P(\gamma)=\Id-\gamma\,m\!\otimes\!n.
\end{equation}
The analysis can be carried out without this assumption, but the formulas
become very lengthy. Secondly, we assume that the total deformation is a shear
of the material along this slip system,
\begin{equation}
\label{shear}
D\phi(t)=\Id+\beta(t)m\!\otimes\!n\qquad\mbox{in }\overline{\Omega}
\end{equation}
for a given scalar function $\beta(t)$. On $\partial\Omega$, this identity must
be imposed by choosing in (\ref{BC}) the appropriate boundary conditions on
$\phi$. Eqn.~(\ref{shear}) states that the deformation at every point in
$\Omega$ follows this prescribed deformation at $\partial\Omega$, i.e.
postulates the validity of the Cauchy-Born rule.

Assuming (\ref{shear}), the mapping $\gamma\mapsto\Wst(\!\Re^tD\phi P(\gamma))$
is convex (see Lemma~\ref{lem1} below). This implies that the constraint
$f_1(\gamma,\kappa)\le0$ is satisfied with equality,
\[ |\gamma-\gamma_0|+\kappa-\kappa_0=0. \]
Therefore we can resolve this condition
and (\ref{Emin2}) can be written as the unconstrained minimisation problem
\begin{eqnarray}
E_\beta(\Re,\gamma) &=& \io\Big[\Wst(\Re^t(\Id\!+\!(\beta\!-\!\gamma)
m\!\otimes\!n))+\Wc(\Ke)+V\big(\kappa_0\!-\!|\gamma\!-\!\gamma^0|\big)\nn\\
\label{Emin3}
 && \hspace*{25pt} +\sigma_Y|\gamma-\gamma^0|\Big]\dx\to\min,\\
{\Re}|_{\partial\Omega} &=& R_D.\nn
\end{eqnarray}

To explicitly compute the solutions to (\ref{Emin3}), we pick
\begin{equation}
\label{Vdef}
V(\kappa):=\rho\kappa^2.
\end{equation}
For $\Wst$, $\Wc$ we make the general ansatz, cf. \cite{Neff06},
\begin{eqnarray}
\label{Wstdef}
\Wst(\Ue) &:=& \mu\|\sym\Ue-\Id\|^2+\mu_c\|\skw\Ue\|^2+\frac\lambda2
|\tr(\Ue-\Id)|^2,\\
\Wc(\Ke) &:=& \mu_2\frac{L_c^{1+p}}{2}(1+\alpha_4L_c^q\|\Ke\|^q)\nn\\
\label{Wcdef0}
&& \times(\alpha_5\|\sym\Ke\|^2+\alpha_6\|\skw\Ke\|^2
+\alpha_7|\tr(\Ke)|^2)^\frac{1+p}{2},
\end{eqnarray}
for positive material parameters $\rho$, $\mu$, $\mu_2$, $\lambda$, $\mu_c$,
constants $L_c>0$, $\alpha_4\ge0$, $\alpha_5>0$, $\alpha_6,\alpha_7\ge0$, $p>0$,
$q\ge0$, and where $\skw A:=0.5(A-A^t)$, 
\[ \|A\|:=\sqrt{\tr(A^tA)} \]
is the Frobenius matrix norm, and $\tr(A):=\sum_i A_{ii}$ the trace operator.
Setting $L_c:=\sqrt2$, $\alpha_5=\alpha_6:=1$, $\alpha_4=\alpha_7:=0$, $p:=1$,
the definition (\ref{Wcdef0}) simplifies to
\[ \Wc(\Ke)=\mu_2\|\Ke\|^2. \]
Since
\begin{eqnarray*}
\|\Ke\|^2 &=& \sum_{l=1}^d\|\Re^t\partial_{x_l}\Re\|^2=\sum_{l=1}^d
\tr\big(\partial_{x_l}\Re^t\Re\Re^t\partial_{x_l}\Re\big)\\
&=& \sum_{l=1}^d\tr\big(\partial_{x_l}\Re^t\partial_{x_l}\Re\big)
=\sum_{l=1}^d\|\partial_{x_l}\Re\|^2,
\end{eqnarray*}
the definition of $\Wc$ simplifies further to
\begin{equation}
\label{Wcdef}
\Wc(\Re):=\mu_2\|\nabla\Re\|^2.
\end{equation}
In two space dimensions, we can benefit from the explicit representation
\begin{equation}
\label{alpha}
\Re=\Re(\alpha)=\left(\begin{array}{rr}\cos\alpha & -\sin\alpha\\ \sin\alpha &
\cos\alpha \end{array}\right),\qquad\alpha\in[0,2\pi).
\end{equation}
Finally, by direct computations,
\begin{eqnarray*}
\|\partial_{x_l}\Re\|^2 &=& \left\|\left(\begin{array}{rr} -\sin\alpha &
-\cos\alpha\\ \cos\alpha & -\sin\alpha \end{array}\right)\partial_{x_l}\alpha
\right\|^2=2|\partial_{x_l}\alpha|^2,\qquad l=1,2,\\
\|\sym\Ue\!-\!\Id\|^2 &=& \|\sym[\Re(-\alpha)(\Id+(\beta\!-\!\gamma)
m\!\otimes\!n)]-\Id\|^2\\
&=& 2(1-\cos\alpha)^2+2(\beta\!-\!\gamma)(1-\cos\alpha)\sin\alpha\\
&& +(\beta\!-\!\gamma)^2\Big(\sin^2\alpha+\half\cos^2\alpha\Big),\\
\|\skw\Ue\|^2 &=& 2\sin^2\alpha+2(\beta\!-\!\gamma)\sin\alpha\cos\alpha
+(\beta\!-\!\gamma)^2\frac{\cos^2\alpha}{2},\\
|\tr(\Ue\!-\!\Id)|^2 &=& 4(1\!-\!\cos\alpha)^2
+4(\beta\!-\!\gamma)(1\!-\!\cos\alpha)\sin\alpha
+(\beta\!-\!\gamma)^2\sin^2\alpha.
\end{eqnarray*}
In summary, using the definitions (\ref{Vdef}), (\ref{Wstdef}), (\ref{Wcdef})
and letting $\kappa_0=\gamma_0=0$ for simplicity, (\ref{Emin3}) becomes
(with $\Re(\alpha_D)=R_D$)
\begin{eqnarray}
E_\beta(\alpha,\gamma) &=& \io\!\Big[2\mu_2|\nabla\alpha|^2+\rho\gamma^2
+\sigma_Y|\gamma|+2(\lambda\!+\!\mu)(1\!-\!\cos\alpha)^2+2\mu_c\sin^2\alpha\nn\\
&& \hspace*{20pt} -2(\gamma\!-\!\beta)\big((\mu_c\!-\!\lambda\!-\!\mu)
\cos\alpha\!+\!\lambda\!+\!\mu\big)\sin\alpha\nn\\
\label{Emin4}
&& \hspace*{20pt} +\frac{(\gamma\!-\!\beta)^2}{2}\big(\mu+(\lambda\!+\!\mu)
\sin^2\alpha+\mu_c\cos^2\alpha\big)\Big]\dx\to\min,\\
\label{BC2}
\alpha|_{\partial\Omega} &=& \alpha_D.
\end{eqnarray}

\begin{Lemma}
\label{lem1}
The mappings $\gamma\mapsto E_\beta(\alpha,\gamma)$ and
$\gamma\mapsto\Wst(\Ue(\alpha,\gamma))$ are convex.
\end{Lemma}
\noindent{\bf Proof} From (\ref{Emin4}) we have
\[ \frac{\partial^2\Wst(\Ue(\alpha,\gamma))}{\partial\gamma^2}
=\mu+(\lambda+\mu)\sin^2\alpha+\mu_c\cos^2\alpha>0. \]
Since $V(\gamma)$ and $\gamma\mapsto\sigma_Y|\gamma|$ are convex, this implies
the convexity of $E_\beta(\alpha,\cdot)$.

\hfill \qed

\vspace*{5mm}
We study (\ref{Emin4}), (\ref{BC2}) for $\alpha\in H^{1,2}(\Omega;\,[0,2\pi))$
and $\beta(t)$, $\gamma\in L^2(\Omega;\,\R)$, where $\beta(t)$ is given and
controls the shear of $\Omega$.
From the direct method in the calculus of variations, since $E$ is coercive and
weakly lower semicontinuous, the existence of solutions $(\alpha,\gamma)$
follows.

\section{Analytic solutions of the 2D shear problem}
\label{secsol}
In this section we study properties of the solution $(\alpha,\gamma)$ of
(\ref{Emin4}) for several choices of the parameters and are especially
interested in cases where transition layers in $\Re$ occur.

\subsection{The limiting case $\mu_2\to\infty$}
\label{case1}
This limiting case corresponds to rigidity in bending. After rescaling,
only the curvature energy $\Wc$ contributes and (\ref{Emin4}) is equivalent to
\[ E(\alpha)=2\io|\nabla\alpha|^2\dx\to\min \]
subject to (\ref{BC2}). The minimiser $\alpha$ solves the classical harmonic
problem
\[ -\triangle\alpha=0\mbox{ in }\Omega,\qquad\alpha|_{\partial\Omega}=\alpha_D.
\]

\subsection{The limiting case $\rho$, $\sigma_Y\searrow0$}
\label{case2}
The limit $\rho\searrow0$, $\sigma_Y\searrow0$ corresponds to an ultra-soft
material where the dislocations do not contribute to the mechanical energy.
Here, Equation~(\ref{Emin4}) simplifies to
\begin{eqnarray}
E_\beta(\alpha,\gamma) \!\!\!&=&\!\!\! \io\!\Big[2\mu_2|\nabla\alpha|^2
+2(\lambda\!+\!\mu)(1\!-\!\cos\alpha)^2+2\mu_c\sin^2\alpha\nn\\
&& \hspace*{17pt} -2(\gamma\!-\!\beta)\big((\mu_c\!-\!\lambda\!-\!\mu)
\cos\alpha\!+\!\lambda\!+\!\mu\big)\sin\alpha\nn\\
\label{min2}
&& \hspace*{17pt} +\frac{(\gamma\!-\!\beta)^2}{2}\big(
\mu\!+\!(\lambda\!+\!\mu)\sin^2\alpha\!+\!\mu_c\cos^2\alpha\big)\Big)\Big]
\dx\to\min.
\end{eqnarray}
Clearly, due to convexity in $\gamma$,
the optimal $\gamma$ follows the given deformation $\beta(t)$,
\[ \gamma_\mathrm{opt}(t)=\beta(t). \]
Computing the Euler-Lagrange equation of (\ref{min2}), we need to find a bounded
solution $\aopt\in[0,2\pi)$ of the semi-linear elliptic partial
differential equation
\begin{equation}
\label{EL}
-\mu_2\triangle\alpha+[\lambda+\mu+(\mu_c-\lambda-\mu)\cos\alpha)]\sin\alpha=0
\mbox{ in }\Omega
\end{equation}
with boundary conditions (\ref{BC2}). This equation has been studied before in
\cite{PZ}. Based on a separation ansatz, special solutions have been found in
\cite{MR93}.

First consider the unphysical case $\mu_2=0$. Then evidently
\begin{equation}
\label{minW}
\aopt\in\{0,\pi\}\mbox{ in }\Omega,
\qquad{\aopt}|_{\partial\Omega}=\alpha_D.
\end{equation}
Since there is no regularising term $|\nabla\alpha|^2$ in $E$,
$\aopt\in\{0,\pi\}$ a.e. in $\Omega$ and $\alpha\in L^2(\Omega;\,[0,2\pi))$
has no additional regularity.

\begin{figure}[hbt]
\unitlength1cm
\begin{picture}(10.0,4.0)
\put(0.2,0){\psfig{figure=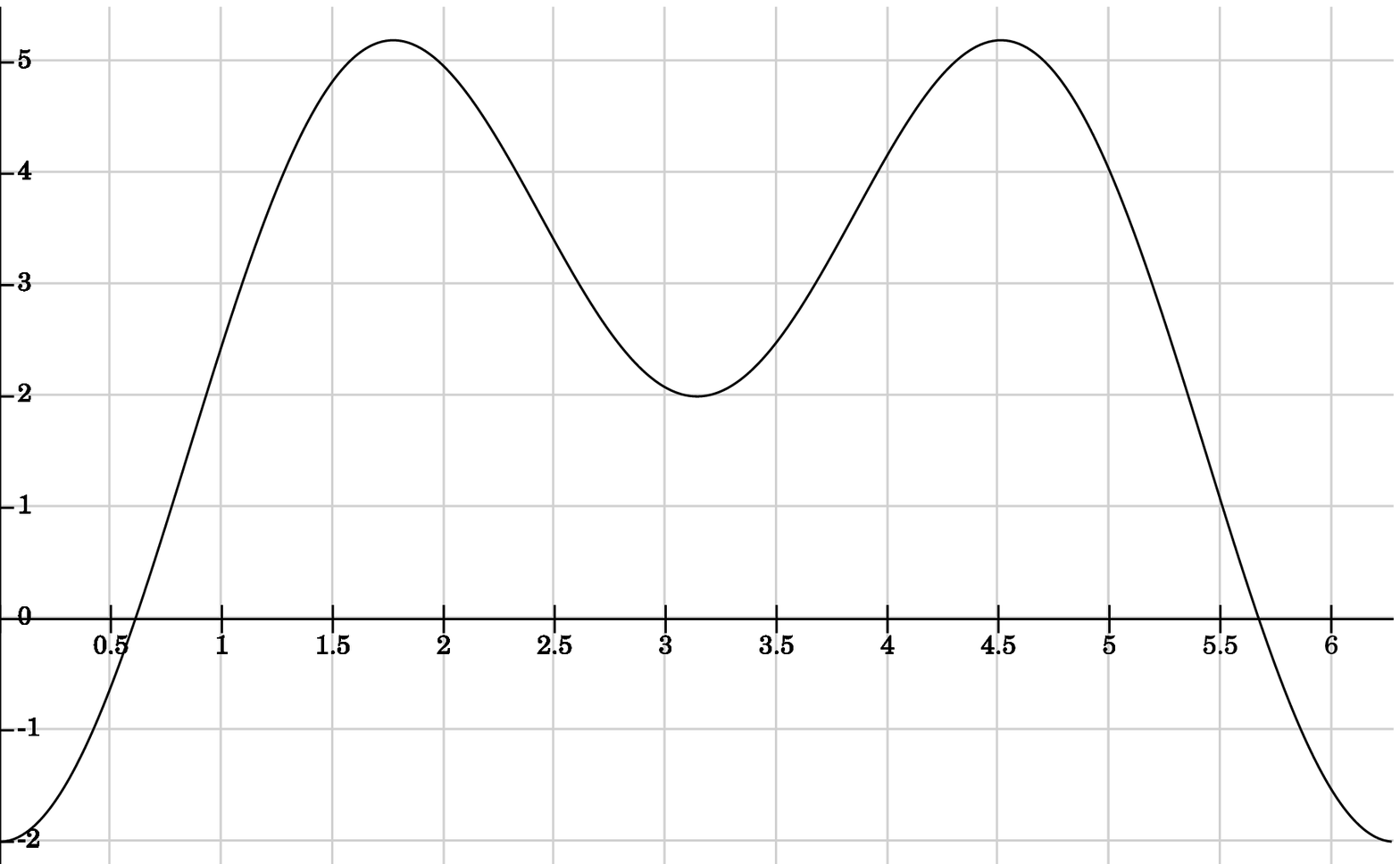,height=40mm}}
\put(7.2,0.05){\psfig{figure=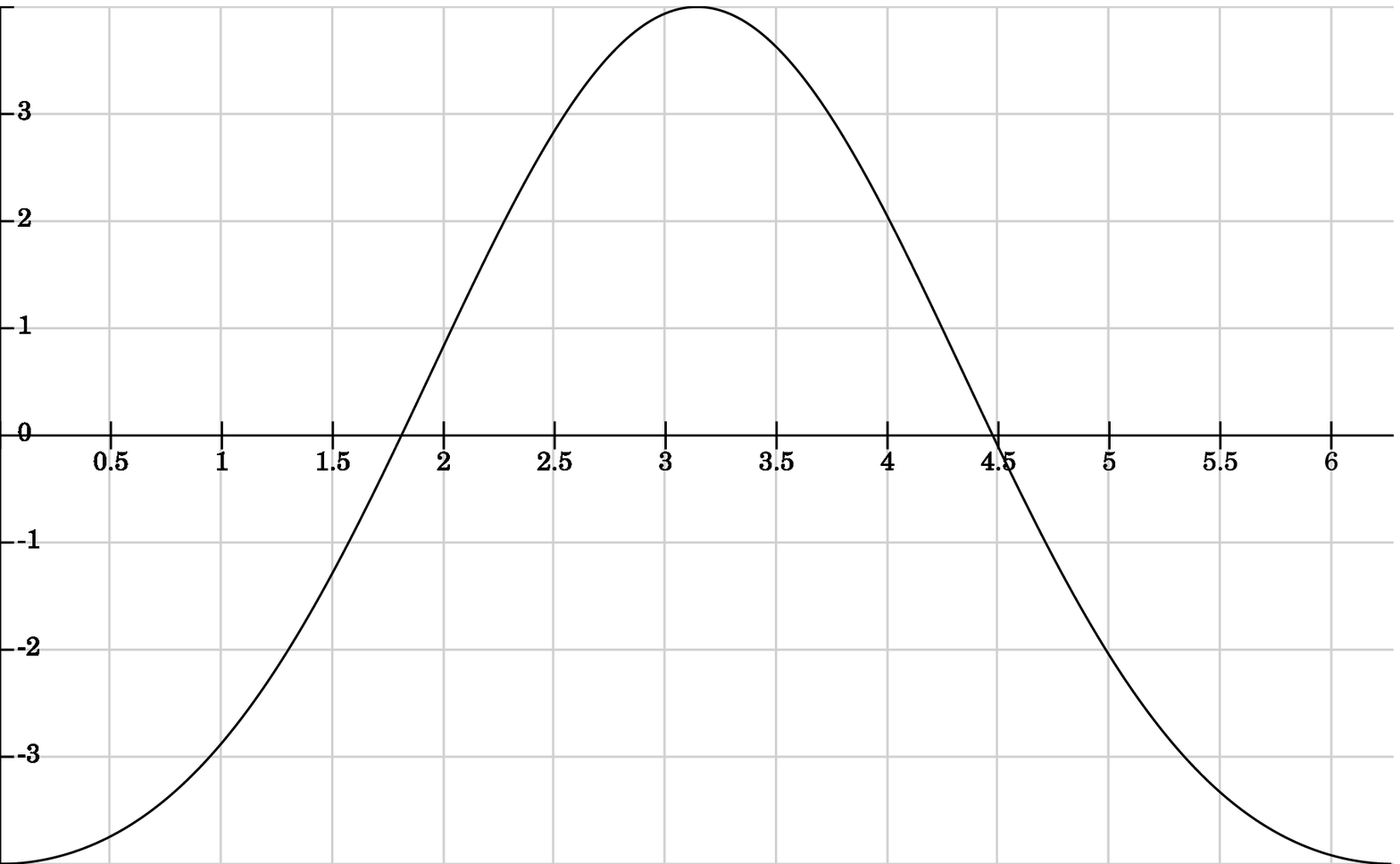,height=40mm}}
\end{picture}
\caption{Plots of $J$ for $\lambda=\mu=1$. Left: $\mu_c=12$.
$J$ has the two local minima $0$ and $\pi$. Right: $\mu_c=1$. $J$ has only one
minimum at $0$.}
\label{fig1}
\end{figure}

When $\mu_2>0$, the solutions to (\ref{EL}) are continuous and stationary
(in time) solutions of the Allen-Cahn equation
\begin{equation}
\label{AC}
\pat\alpha=\mu_2\triangle\alpha-J'(\alpha)
\end{equation}
with the nonlinearity
\begin{equation}
\label{Jdef}
J(\alpha):=-(\lambda+\mu)\cos\alpha+\frac{\mu_c-\lambda-\mu}2\sin^2\alpha.
\end{equation}
$J$ has a local minimum at $0$.
For
\begin{equation}
\label{mccond}
\mu_c>2(\lambda+\mu),
\end{equation}
there is exactly one other local minimum at $\alpha=\pi$
as illustrated in Fig.~\ref{fig1}.

The properties of the solution to (\ref{AC}) when (\ref{mccond}) holds are
well known: $\alpha$ tends pointwise to one of the minimisers of $J$ (here
$0$ and $\pi$), but transition layers of width $\sqrt{\mu_2}$ and slope
$C/\sqrt{\mu_2}$ occur. For small $\mu_2$, the interface moves by its negative
mean curvature.
For $t\to\infty$, $\alpha$ becomes stationary and when periodic boundary
conditions are applied, $\alpha\equiv0$ or $\alpha\equiv\pi$ in $\Omega$.
If Dirichlet boundary conditions (\ref{BC2}) are imposed, a sufficient
condition for the presence of transition layers in the stationary limit
$t\to\infty$ is
\begin{equation}
\label{TL1}
\Big(0\!\le\!\alpha_D\!<\!\frac\pi2\Big)\mbox{ or }\Big(
\frac{3\pi}2\!<\!\alpha_D\!<\!2\pi\Big)\mbox{ in }D_1,\qquad
\frac\pi2\!<\!\alpha_D\!<\!\frac{3\pi}2 \mbox{ in }D_2,
\end{equation}
where $D_1$, $D_2\subset\partial\Omega$ are disjoint sets with positive
Hausdorff measure.

\subsection{The full problem in the elastic regime}
\label{case3}
Now we investigate a case complementary to the one previously studied
where the material is so hard that no plastic flow occurs. We assume again
$\gamma_0=\kappa_0=0$. Computing the first variation of
$E_\beta$ given by (\ref{Emin4}) w.r.t. $\gamma$, since
\[ \partial^\mathrm{sub}|\gamma|=\left\{\begin{array}{ll} +1, &\mbox{if }
\gamma>0,\\ -1, &\mbox{if } \gamma<0,\\
\;\![-1,+1], &\mbox{if }\gamma=0, \end{array}\right. \]
we obtain that no plastic flow occurs (i.e. $\gamma=\gamma_0$) if
\begin{equation}
\label{s1}
-2((\mu_c\!-\!\lambda\!-\!\mu)\cos\alpha\!+\!\lambda\!+\!\mu)\sin\alpha
-\beta(\mu\!+\!(\lambda\!+\!\mu)\sin^2\alpha\!+\!\mu_c\cos^2\alpha)\in
[-\sigma_Y,\sigma_Y].
\end{equation}
A sufficient condition independent of $\alpha$ for (\ref{s1}) is
\begin{equation}
\label{s2}
2\max\{2(\lambda+\mu)-\mu_c,\mu_c\}+|\beta(t)|\big(
\lambda+2\mu+\mu_c\big)\le\sigma_Y.
\end{equation}
For fixed $\lambda$, $\mu$ and $\mu_c$, Condition~(\ref{s2}) is satisfied if
the yield stress $\sigma_Y$ of the deformed material is large enough and then,
(\ref{s2}) turns into a smallness condition on $|\beta(t)|$.

The rotations $\Re$ are determined as solutions of the minimisation problem
\begin{eqnarray*}
E_\beta(\alpha) &:=& \io\Big[2\mu_2|\nabla\alpha|^2+2(\lambda\!+\!\mu)
(1\!-\!\cos\alpha)^2+2\mu_c\sin^2\alpha\\
&&\hspace*{25pt} +2\beta(t)\big((\mu_c\!-\!\lambda\!-\!\mu)\cos
\alpha\!+\!\lambda\!+\!\mu)\sin\alpha\\
&&\hspace*{25pt} +\frac{\beta(t)^2}{2}\big(\mu\!+\!(\lambda\!+\!\mu)\sin^2
\alpha\!+\!\mu_c\cos^2\alpha\big)\Big]\dx\to\min
\end{eqnarray*}
subject to (\ref{BC2}).

When computing the Euler-Lagrange equation, we thus need to
find $\alpha\in L^2(\Omega;\,[0,2\pi))$ which solves
\begin{eqnarray*}
-\mu_2\triangle\alpha \!\!\!&+&\!\!\! \Big[\lambda\!+\!\mu
+\Big(\mu_c\!-\!\lambda\!-\!\mu)
\Big(1\!-\!\frac{\beta(t)^2}{4}\Big)\!\cos\alpha\Big]\sin\alpha\\
&+& \frac{\beta(t)}{2}\Big((\lambda\!+\!\mu)\cos\alpha\!+\!\big(\mu_c
\!-\!\lambda\!-\!\mu\big)\big(\cos^2\alpha\!-\!\sin^2\alpha\big)\Big)=0\quad
\mbox{in }\Omega.
\end{eqnarray*}
(By standard elliptic regularity theory, then 
$\alpha\in H^2(\Omega;\,[0,2\pi))\;$).

This time it remains to find stationary solutions to the Allen-Cahn equation
(\ref{AC}) with
\begin{eqnarray}
J_\beta(\alpha) &:=& -\Big(\lambda\!+\!\mu+\beta(t)
\frac{\lambda\!+\!\mu\!-\!\mu_c}{2}\sin\alpha\Big)\cos\alpha
+\frac{\beta(t)}{2}(\lambda\!+\!\mu)\sin\alpha\nn\\
\label{J2def}
&& +\, \frac{\mu_c\!-\!\lambda\!-\!\mu}{2}\Big(1\!-\!\frac{\beta(t)^2}{4}\Big)
\sin^2\alpha.
\end{eqnarray}
For $\beta=0$, $J_\beta$ is identical to (\ref{Jdef}) and the
partitioning of $\Omega$ into subsets with $\alpha\sim0$ and $\alpha\sim\pi$
takes place if (\ref{mccond}) holds as discussed in the previous subsection.
When $\beta\not=0$, the discussion is more elaborate as $J_\beta$ depends on
$2$ independent parameters and $\alpha$. Since $J_\beta(\alpha)$ is
particularly simple for $\beta=\pm2$, let $T>0$ be a stop time and $\beta(t)$
be any continuous curve with $\beta(0)=-2$, $\beta(T)=+2$
that fulfils (\ref{s2}) uniformly in $0\le t\le T$.

\begin{figure}[hbt]
\unitlength1cm
\begin{picture}(10.0,5.5)
\put(1.8,-0.5){\psfig{figure=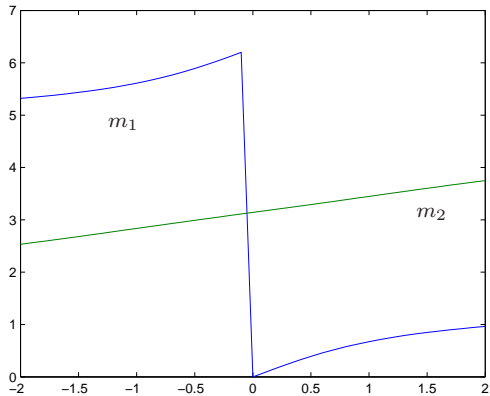,height=60mm}}
\put(4.0,3.5){\scriptsize$m_1$}
\put(8.1,2.3){\scriptsize$m_2$}
\end{picture}
\caption{The two local minima $m_1(\beta)$, $m_2(\beta)$ of $J_\beta$ as a
function of $\beta\in[-2,2]$ for $\mu_c=6\mu$, $\lambda=\mu$.}
\label{fig2}
\end{figure}

Fig.~\ref{fig2} displays a typical plot of the two distinct local minima
$m_1$, $m_2$ of $J_\beta$ for $\beta\in[-2,2]$ computed with a one-dimensional
Newton method, here for the special choice $\mu_c=6\mu$, $\lambda=\mu$ which
satisfies (\ref{mccond}).

Like the condition (\ref{TL1}), we can derive sufficient conditions on
$\alpha_D$ for the formation of boundary layers. A strong time-independent
condition is
\begin{eqnarray}
\Big(0\!\le\!\alpha_D\!<\!M_1(2)\Big)\mbox{ or }
\Big(M_2(-2)\!<\alpha_D\!<\!2\pi\Big)\mbox{ in }D_1,\nn\\
\label{TL2}
M_1(-2)<\alpha_D<M_2(2)\mbox{ in }D_2.
\end{eqnarray}
where $D_1$, $D_2\subset\partial\Omega$ are disjoint sets with positive
Hausdorff measure; $M_1(t)<M_2(t)$ are the local maxima of $J_\beta$, and it
was used that $M_1(t)$ is strictly increasing,
$M_2(t)$ strictly decreasing in time.

\section{Discussion and concluding remarks}
\label{secdiscussion}
In this article we computed analytically solutions of a 2D shear problem
within the framework of finite-strain Cosserat plasticity.

As main result of our investigations, we considered two complementary cases of
shear in 2D, one of an ultra-soft material where plasticity occurs during the
entire deformation process, and another case of a hard material where the
applied loads are not large enough to initiate plastic flow.
In both examples it was shown that the micro-rotations $\Re$ are stationary
solutions of the Allen-Cahn equation (\ref{AC}) which may result in a
partitioning of the material into cells due to the occurrence of transition
layers in $\Re$. If $\sqrt{\mu_2}$ is small (compared to $|\Omega|$), the
transition layers are steep and $\Re$ is approximately constant in each cell
with a value determined by one of the local minimisers of a functional $J$ that
depends on the parameters $\lambda+\mu$, $(\mu_c-\lambda-\mu)/2$ and the
applied deformations $\beta(t)$, cf. Eqn.~(\ref{J2def}).

The simple example in Subsection~\ref{case1} demonstrates that this partitioning
is restricted to a certain parameter range of $\mu_2$ and does not always occur.
In addition, like the celebrated Cauchy-Born rule, the values of $\Re$
at $\partial\Omega$ determine the behaviour in the interior, especially
whether and where transition layers occur.

Yet, in many cases the analysis suggests a subdivision mechanism of $\Omega$
caused by deforming the solid that may lead to an explanation why grains and
subgrains form. Along this line, it is essential to work out the precise
relation between local orientation $\Re$ and orientation of the subgrain.
A first investigation in 3D using finite-element computations is
done in \cite{Forest00}. In this work, it was assumed that both coincide.

Since the analysed model relies on simplifying assumptions, it is natural to ask
whether the proposed mechanism carries over to the full three-dimensional
setup, to general deformations, and to more realistic
dislocation models. Another very important
question left to future work is the study of $\ve$-minimisers, characterised by
\[ E(\phi_\ve,{\Re}_\ve,{\Fp}_\ve,\kappa_\ve)\le m+\ve, \]
where $m$ is the absolute minimal value of $E$ and $\ve>0$ a small number.
This notion goes along with the insight that most of the time, the solutions
observed in experiments do not reach a global minimum. If an almost optimal
micro-rotation ${\Re}_\ve$ solves the time-dependent Allen-Cahn equation
(\ref{AC}) for large but finite time $t$, this may pave the way for quantitative
studies since the coarsening laws of the Allen-Cahn equation are well known.

\section*{Acknowledgements}
TB acknowledges the support by the Hausdorff institute of Mathematics, Bonn and
the support by the German Research Community (DFG) through grant BL 512 4/1.

\end{document}